\documentclass[cup5b]{caps}
\usepackage{graphicx}
\usepackage{amssymb}
\usepackage{ociwsymp4e}  %Use this for contributed papers.

\HeadText{E.~Pancino}  %Enter author name; if three authors, list all three;
\begin{document}

\pagenumbering{arabic}

\author[]{E. PANCINO\\INAF -- Bologna Observatory, Italy}

\chapter{Metal Rich Stars in $\omega$ Centauri}

%%%%%%%%%%%%%%%%%%%%%%%%%%%%%%%%%%%%%%%%%%%%%%%%%%%%%%%%%%%%%%%%%%%%%%
%%% ABSTRACT %%%%%%%%%%%%%%%%%%%%%%%%%%%%%%%%%%%%%%%%%%%%%%%%%%%%%%%%%
%%%%%%%%%%%%%%%%%%%%%%%%%%%%%%%%%%%%%%%%%%%%%%%%%%%%%%%%%%%%%%%%%%%%%%

\begin{abstract}

I present some preliminary results from a high-resolution spectroscopic
observing campaign conducted with UVES (Ultraviolet Visual Echelle
Spectrograph) at the ESO VLT (Very Large Telescope), devoted to the
study of the newly discovered, metal-rich red giant branch in
$\omega$~Centauri (the RGB-a). In particular, I derive and discuss
accurate abundances of iron-peak elements, $\alpha$-elements and
$s$-process elements. The main results discussed in this contribution
are: {\it (i)} the RGB-a is the most metal-rich component of the RGB
stellar mix in $\omega$~Cen, with [Fe/H]=$-0.62\pm0.06$; {\it (ii)} the
RGB-a has a lower $\alpha$-enhancement compared to the other red
giants, possibly due to SNe~type~Ia pollution and {\it (iii)} the
$s$-process elements overabundance of the RGB-a is similar to that of
the other red giants in $\omega$~Cen, and unusually high for globular
cluster stars, due to pollution by low mass Asymptotic Giant Branch
(AGB) stars.

\end{abstract}

%%%%%%%%%%%%%%%%%%%%%%%%%%%%%%%%%%%%%%%%%%%%%%%%%%%%%%%%%%%%%%%%%%%%%%
%%% INTRODUZIONE %%%%%%%%%%%%%%%%%%%%%%%%%%%%%%%%%%%%%%%%%%%%%%%%%%%%%
%%%%%%%%%%%%%%%%%%%%%%%%%%%%%%%%%%%%%%%%%%%%%%%%%%%%%%%%%%%%%%%%%%%%%%

\section{Introduction}

\begin{figure}
\centering
%\plotone{pancino.fig02.ps}
\includegraphics[width=9cm]{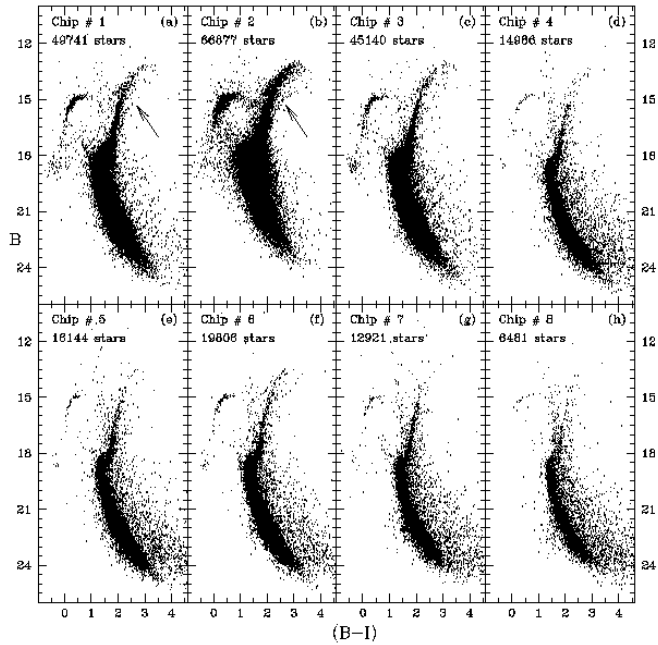}

\caption{The $B$,($B-I$) color magnitude diagrams (CMD) for
$\omega$~Cen (Pancino et al. 2000). Each CMD is obtained from one of
the CCD chips in the WFI mosaic. The cluster center is located on Chip
\#2. The arrows indicate the anomalous RGB (RGB-a), which is only
visible in the center and, to a lesser extent, in the neighbouring
CCDs.}

\label{wfi}
\end{figure}

$\omega$~Centauri (NGC 5139) is the most massive and bright system
among the Galactic Globular Clusters (GGC). However, its most
astonishing characteristic is the chemical inhomogeneity: $\omega$~Cen
is the only GGC that shows clear and undisputed variations not only in
the light elements abundances, but also in its overall metallicity.
From this point of view, $\omega$~Cen could be considered a ``bridge''
between the common globulars, which are unable to retain the gas
ejected by their former massive stars, and the dwarf galaxies, which
are the least massive self-enriching stellar systems known. It is
interesting to note that those dwarf spheroidal galaxies (dSph) that
are less luminous than $\omega$~Cen (e.g. Ursa Minor, Draco and Carina)
show modest abundance spreads with respect to this cluster.

The metallicity distribution of $\omega$~Cen red giant branch (RGB)
stars was derived by means of low resolution spectroscopy (see for
example Norris et al. 1996 or Suntzeff \& Kraft 1996), showing its main
peak at [Fe/H]=$-1.6$ with a long extended tail to higher metallicities
and a possible secondary peak at about [Fe/H]$\simeq-1.2$. Recently,
wide field photometric studies of $\omega$~Cen (Lee et al. 1999 and
Pancino et al. 2000) have revealed the complex structure of the RGB of
$\omega$~Cen. In particular, the presence of an additional well
detached sequence, previously unknown, to the red side of the RGB
(Figure~\ref{wfi}) has been discovered. According to Pancino et al.
(2000), three main sub-populations can be identified in the RGB
distribution of $\omega$~Cen: {\it (i)} the main, metal-poor
population (RGB-MP), with [Fe/H]=$-1.6$ and comprising $\sim70\%$ of
the RGB population; {\it (ii)} the RGB-MInt, comprising the secondary
peak at [Fe/H]$\simeq-1.2$ together with the long tail extending to
higher metallicity, comprising $\sim25\%$ of the RGB; {\it (iii)} the
additional RGB-a recently identified, that appears to be the most
metal-rich component in $\omega$~Cen ([Fe/H]=$-0.6$, Pancino et al.
2002) and comprises only $\sim5\%$ of the RGB population.

We have started a long term project devoted to the photometric and
spectroscopic study of the various sub-populations in $\omega$~Cen, 
which is described in Ferraro et al. (2002) and in Pancino et al.
(2003). I present here the first, preliminary results from the optical,
high-resolution survey conducted with UVES mounted at the ESO VLT,
which has been undertaken to investigate on the chemical properties of
the RGB-a population, which was never studied with high-resolution
spectroscopy before.

%%%%%%%%%%%%%%%%%%%%%%%%%%%%%%%%%%%%%%%%%%%%%%%%%%%%%%%%%%%%%%%%%%%%%%
%%% DATI %%%%%%%%%%%%%%%%%%%%%%%%%%%%%%%%%%%%%%%%%%%%%%%%%%%%%%%%%%%%%
%%%%%%%%%%%%%%%%%%%%%%%%%%%%%%%%%%%%%%%%%%%%%%%%%%%%%%%%%%%%%%%%%%%%%%

\section{Observational Material}

\begin{figure}
\centering
%\plotone{pancino.fig01.ps}
\includegraphics[width=8cm]{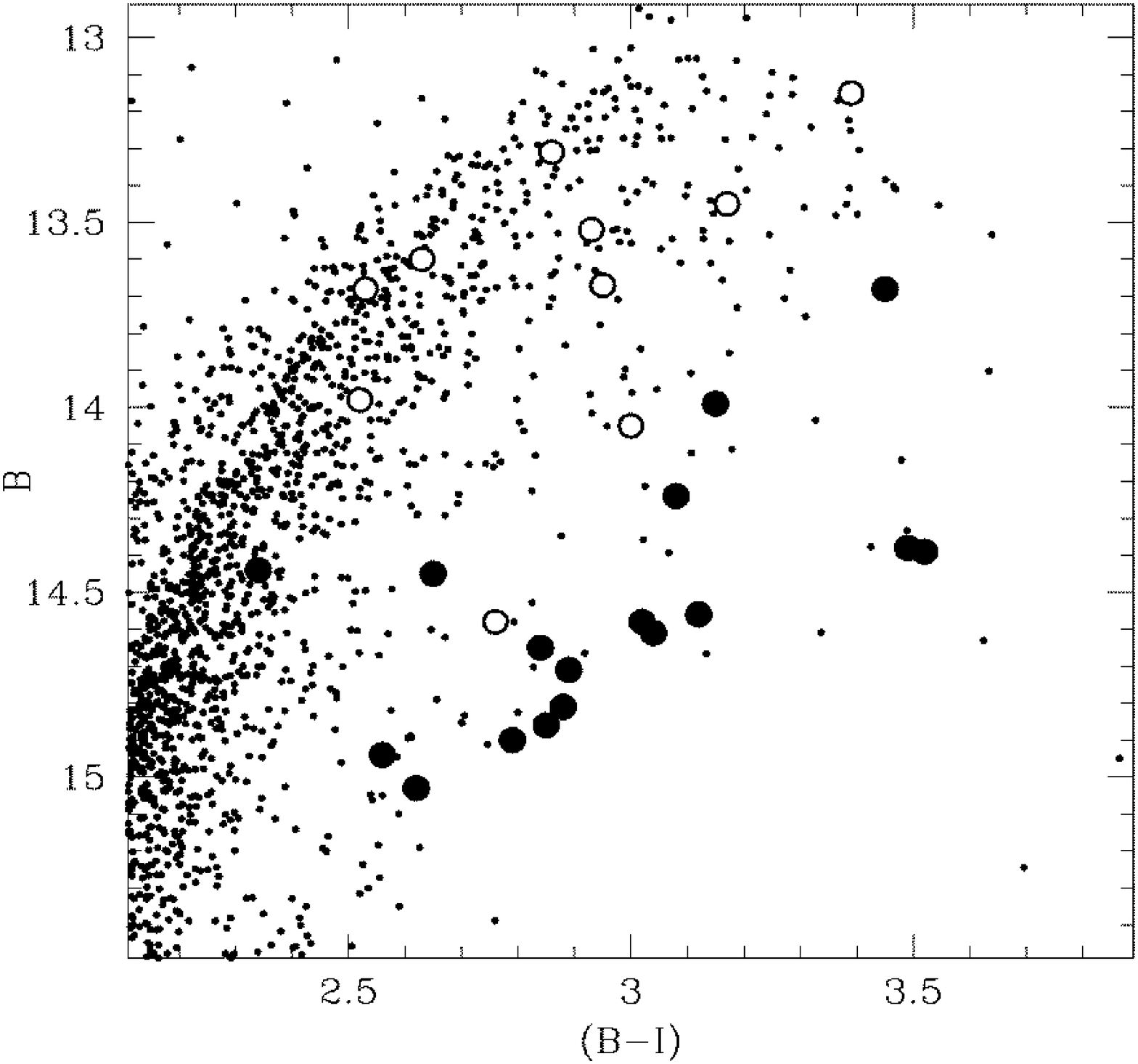}

\caption{The RGB of $\omega$~Cen is plotted as small dots, where the
targets of the UVES survey are marked with large circles: {\it empty
circles} for the RGB-MP and RGB-MInt stars that have not been analyzed
yet, {\it filled circles} for the RGB-a and RGB-MInt stars that are
presented in this contribution.}

\label{targets}
\end{figure}

The principal aim of the observations was to cover the RGB-a along its
full extension in a homogeneous way while having, at the same time, a
good number of RGB-MInt and RGB-MP giants in common with previous
high-resolution abundance studies of $\omega$~Cen. In particular, the
attention has been concentrated so far on the most metal-rich stars in
the sample, mostly belonging to the RGB-a and RGB-MInt populations
(Figure~\ref{targets}). 

Observations were conducted during three different observing runs with
UVES at the ESO VLT in Paranal, Chile. All the spectra have
high-resolution ($R\simeq45,000$) and high signal to noise ratios (see
Table~\ref{logs}). The first run (Run~A) was carried out in June 2000,
as a backup while the main targets were not visible, and six stars were
observed covering a wide wavelength range (3500--9000\AA, with gaps
around 5000\AA\  and 7000\AA). A preliminary analysis of these stars
has been already published by Pancino et al. (2002). The rest of the
sample has been observed in April 2001 (Run~B), with a slightly
different setup, covering the range 3500--7000\AA, with two gaps around
4600\AA\ and 5800\AA. Two additional stars have been added during the
last observing run in March 2002 (Run~C), using the same setup as in
Run~B. 

A summary of the stars properties is presented in Table~\ref{logs}.
Colors and magnitudes are from the wide field CCD $B$ and $I$
photometry published by Pancino et al. (2000) and from some additional
$V$ band material observed with the WFI (Wide Field Imager) at the ESO
2.2~m telescope in La Silla, Chile. The colors have been dereddened
using $E(B-V)=0.11$ from Lub (2002).

\begin{table}

\caption{Target information for the stars analyzed here. The columns
contain: {\it (1)-(2)} the star name from Pancino et al. (2000) and
Woolley et al. (1966); {\it (3)-(5)} magnitudes and dereddened colors
(see text); {\it (6)} observing run; {\it (7)} S/N ratio obtained in
the blue ($\lambda<5000$\AA) and red ($\lambda>5000$\AA) part of the
spectra, respectively and {\it (8)} the sub-population to which the
star belongs.}

\begin{tabular}{cccccccl}
\hline \hline
WFI & ROA & V &  (B--V)$_o$ & (V--I)$_o$ & Run & S/N & Pop \\
{\it (1)} & {\it (2)} & {\it (3)} & {\it (4)} & {\it (5)} & {\it (6)} & 
{\it (7)} & {\it (8)} \\
\hline
126361 & 179 & 12.10 & 1.46 & 1.72 &C & 20/110 & RGB-MInt  \\
139267 & --- & 13.09 & 1.17 & 1.10 &B & 20/90  & RGB-MInt  \\
140419 & --- & 13.49 & 1.22 & 1.19 &B & 25/100 & RGB-a \\
221132 & 300 & 12.71 & 1.48 & 1.63 &A & 30/130 & RGB-a \\
221376 & 500 & 13.12 & 1.30 & 1.36 &B & 30/110 & RGB-a \\ 
221647 & 523 & 13.39 & 1.13 & 1.27 &B & 30/110 & RGB-a \\
222068 & --- & 12.95 & 1.42 & 1.32 &A & 30/110 & RGB-a \\
222679 & --- & 13.26 & 1.26 & 1.25 &A & 25/90  & RGB-a \\
249639 & 517 & 13.10 & 1.19 & 1.35 &B & 25/110 & RGB-a \\
263340 & --- & 13.62 & 1.13 & 1.05 &B & 25/100 & RGB-a \\
305654 & --- & 13.38 & 1.29 & 1.18 &B & 40/110 & RGB-a \\
321293 & --- & 13.69 & 1.15 & 1.09 &B & 30/100 & RGB-a \\
329557 & 447 & 12.85 & 1.35 & 1.79 &B & 30/150 & RGB-a \\
617829 & 371 & 12.71 & 1.34 & 1.36 &A & 40/120 & RGB-MInt  \\
618774 & --- & 13.22 & 1.24 & 1.22 &B & 25/90  & RGB-a \\
618854 & --- & 13.26 & 1.00 & 0.97 &A & 40/90  & RGB-MInt  \\
619210 & 211 & 12.43 & 1.37 & 1.40 &A & 30/140 & RGB-MInt  \\
\hline \hline
\end{tabular}
\label{logs}
\end{table}

The monodimensional spectra were extracted with the UVES pipeline
(Ballester et al. 2000), which includes bias-subtraction, inter-order
background light subtraction, optimal extraction with cosmic ray
rejection, flat-field correction, wavelength calibration, rebinning and
final merging of all overlapping orders. The spectra were then
normalized to the continuum and corrected for telluric absorption bands
using standard tasks within the IRAF\footnote{Image Reduction and
Analysis Facility. IRAF is distributed by the National Optical
Astronomy Observatories, which is operated by the association of
Universities for Reasearch in Astronomy, Inc., under contract with the
National Science Foundation} package {\it noao.onedspec}. To this aim,
a few hot stars have been chosen from the Bright Star Catalog (Hoffleit
\& Jaschek 1991) and observed as telluric standards at least once each
night, at an airmass similar to that of the targets. All of the program
stars presented here are radial velocity members of $\omega$~Cen.

%%%%%%%%%%%%%%%%%%%%%%%%%%%%%%%%%%%%%%%%%%%%%%%%%%%%%%%%%%%%%%%%%%%%%%
%%% ANALISI DELLE ABBONDANZE %%%%%%%%%%%%%%%%%%%%%%%%%%%%%%%%%%%%%%%%%
%%%%%%%%%%%%%%%%%%%%%%%%%%%%%%%%%%%%%%%%%%%%%%%%%%%%%%%%%%%%%%%%%%%%%%

\section{Abundance analysis}

\begin{table}
\centering

\caption{Best models for the stars analyzed so far. The columns
contain: {\it (1)-(2)} star names as in Table~\ref{logs}; {\it (3)-(6)}
the temperature, gravity, microturbulence and metallicity of the
adopted best model and {\it (7)} the resulting iron abundance.}

\begin{tabular}{lllllll}
\hline \hline
WFI & ROA & $T_{eff}$ & $\log~g$ & $v_t$ & [M/H] & [Fe/H] \\
{\it (1)} & {\it (2)} & {\it (3)} & {\it (4)} & {\it (5)} & {\it (6)} 
& {\it (7)} \\
\hline
126361 & 179 & 3900 & 0.6 & 1.8 & -1.0 & -1.34 \\ 
139267 &  -  & 4300 & 1.4 & 1.5 & -0.6 & -0.79 \\ 
140419 &  -  & 4200 & 1.5 & 1.4 & -0.6 & -0.68 \\
221132 & 300 & 4000 & 0.7 & 1.7 & -0.6 & -0.90 \\ 
221376 & 500 & 4100 & 1.5 & 1.2 & -0.6 & -0.52 \\ 
221647 & 523 & 4200 & 1.6 & 1.4 & -0.6 & -0.65 \\ 
222068 &  -  & 4000 & 1.1 & 1.3 & -0.6 & -0.64 \\ 
222679 &  -  & 4200 & 1.6 & 1.3 & -0.6 & -0.53 \\ 
249639 & 517 & 4100 & 1.3 & 1.2 & -0.6 & -0.53 \\ 
263340 &  -  & 4400 & 1.9 & 1.2 & -0.6 & -0.63 \\ 
305654 &  -  & 4200 & 1.6 & 1.4 & -0.6 & -0.65 \\ 
321293 &  -  & 4300 & 1.6 & 1.4 & -0.6 & -0.71 \\ 
329557 & 447 & 4000 & 0.9 & 1.6 & -0.6 & -0.91 \\ 
617829 & 371 & 4000 & 0.8 & 1.7 & -0.6 & -0.93 \\ 
618774 &  -  & 4200 & 1.6 & 1.5 & -0.6 & -0.64 \\ 
618854 &  -  & 4600 & 1.2 & 1.5 & -1.0 & -1.17 \\ 
619210 & 211 & 4000 & 0.7 & 2.1 & -1.0 & -1.08 \\ 
\hline \hline
\end{tabular}

\label{best} 
\end{table}

The abundance analysis has been performed using a fairly standard
spectral synthesis method, similar to the one described in Pancino et
al. (2002), but with more updated software, atomic data and models (see
below). Since we wanted to have all measurements in a common scale, we
re-analyzed the six stars of Run~A together with Runs~B and C. 

We selected a more extended input line list, including many lines for a
set of different elements: {\it (i)} $\alpha$-elements like O, Mg, Si,
Ca and Ti; {\it (ii)} some iron-peak elements like V, Sc, Zn, Cr, Co,
Ni; {\it (iii)} some of the best studied $s$-process elements like Yr,
Zr, Ba, Ce, for a total of $\sim500$ lines. Atomic data from these
lines have been taken from the VALD\footnote{In Pancino et al. (2002)
we used the NIST database instead, which has different oscillator
strengths for many lines. In particular, the difference in [Ca/Fe]
between Pancino et al. (2002) and here is to ascribe entirely to the
choice of log~gf values. Relative abundances among different stars are,
of course, unchanged.} (Vienna Atomic Lines Database -- Kupka et al.
1999).

Equivalent Widths (EW) have been measured by gaussian fitting lines in
the monodimensional spectra by means of an automatic FORTRAN routine
(DAOSPEC, Stetson \& Pancino, in preparation). The resulting EW are in
perfect agreement with the IRAF measurements used in Pancino et al.
(2002), at the $\sim1\%$ level.

\subsection{Search for the Best Model}

First estimates of the effective temperature $T_{eff}$ and surface
gravity $\log g$ have been obtained from the WFI photometry, assuming
$E(B-V)=0.11$ (Lub 2002) and $(m-M)_V=13.97$ (Harris 1996), and using
the color temperature calibrations of Montegriffo (1998) and Alonso et
al. (1999). Estimates of the microturbulent velocity $v_t$ and of the
model metallicity [M/H] have been derived from the color magnitude
diagrams, the curves of growth of iron and previous literature
abundance studies, when available.

Finally, we used the recently updated, expanded models from Plez (1997,
2000 private communication), that span a wider range in gravities,
$\alpha$-enhancements, temperatures and metallicities, to derive
abundances of Fe~I and Fe~II for $\sim9000$ different models for each
star, with parameters close to the first estimates.

The best model has been chosen by imposing simultaneously the four
following conditions: {\it (i)} excitation equilibrium, i.e. Fe lines
with different $\chi_{ex}$ must give the same abundance; {\it (ii)}
lines with different EW must give the same abundance; {\it (iii)}
ionization equilibrium, i.e. lines from Fe~I and Fe~II must give the
same abundance, within the uncertainties and {\it (iv)} lines at
different wavelengths should give the same abundance.

The adopted best models for the stars presented here are shown in
Table~\ref{best} and have been used to measure abundances of the other
elements. An accurate error estimate has not been computed yet, so in
what follows we will attach a representative $0.15$~dex errobar to
every abundance determination. To place the present results in a
broader context, Table~\ref{literature} compares the results for
ROA~179 and ROA~371 with previous literature determinations.

\begin{table}
\centering

\caption{Literature comparisons.}

\begin{tabular}{lllllll}
\hline \hline
WFI & ROA & $T_{eff}$ & $\log~g$ & $v_t$ & [Fe/H] & Reference\\
\hline
126361 & 179 & {\it 3900} & {\it 0.6} & {\it 1.8} & {\it -1.34} & {\it Here} \\ 
       &     &   3850     &    0.5    &     1.5   &     -1.10   & Norris \& Da Costa (1995) \\
       &     &   3850     &    0.5    &     1.0   &     -1.62   & Brown et al. (1991) \\
\hline
617829 & 371 & {\it 4000} & {\it 0.8} & {\it 1.7} & {\it -0.93} & {\it Here} \\ 
       &     &   4000     &    0.9    &     1.6   &     -0.79   & Norris \& Da Costa (1995) \\
       &     &   4000     &    0.9    &     2.2   &     -1.00   & Vanture et al. (1994) \\
       &     &   4000     &    0.9    &     1.5   &     -0.90   & Brown et al. (1991) \\
       &     &   4000     &    0.9    &     2.5   &     -1.37   & Paltoglou \& Norris (1989) \\
\hline \hline
\end{tabular}

\label{literature} 
\end{table}

%%%%%%%%%%%%%%%%%%%%%%%%%%%%%%%%%%%%%%%%%%%%%%%%%%%%%%%%%%%%%%%%%%%%%%
%%% RISULTATI %%%%%%%%%%%%%%%%%%%%%%%%%%%%%%%%%%%%%%%%%%%%%%%%%%%%%%%%
%%%%%%%%%%%%%%%%%%%%%%%%%%%%%%%%%%%%%%%%%%%%%%%%%%%%%%%%%%%%%%%%%%%%%%

\section{Results}

The results presented here are preliminary in the sense that they
concern a sub-sample of the stars observed, of the measurable lines of
each species, of the available atomic species, and of the observed
spectral range. A complete analysis of the dataset should thus add more
information and statistics, but we do not expect it to alter
significantly the abundances obtained up to now. 

A first result concerns the average metallicity of the RGB-a, which
turns out to be [Fe/H]$=-0.62\pm0.06$, perfectly compatible with what
obtained by Pancino et al. (2002), with of course a lower error due to
to the higher number of stars. Moreover, as expected, the detailed
chemical abundance ratios of the iron-peak elements analyzed appear to
be roughly solar at all [Fe/H] values, although some of them, like V
and Sc, appear slightly higher than solar: a detailed spectral
synthesis of each line, including hyperfine structure may lower these
values (see e.g. Smith 2002). 

More interesting is the situation for the $\alpha$-elements and the
$s$-process elements.

\subsection{The $\alpha$-elements}

\begin{figure}
\centering
%\rotatebox{270}{\plotone{pancino.fig03.ps}}
\includegraphics[width=8cm,angle=270]{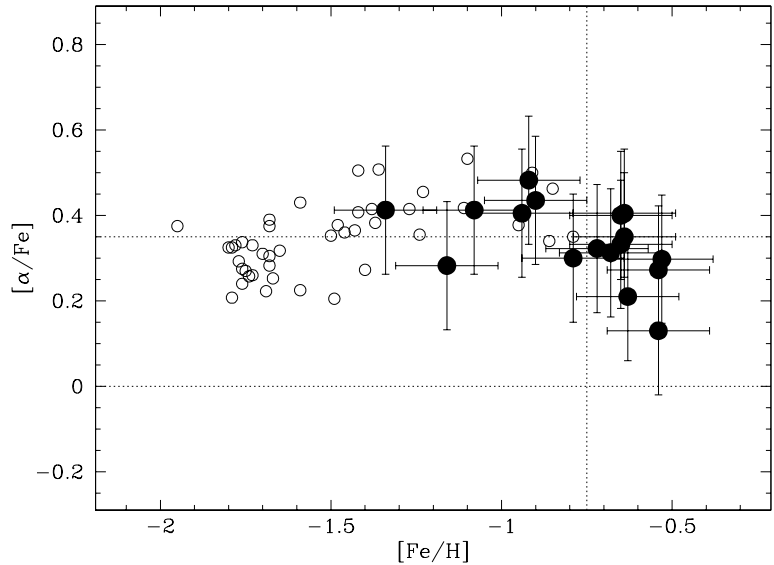}

\caption{The $\alpha$-enhancement is plotted as a function of [Fe/H].
The horizontal dotted lines mark the solar and +0.3 [$\alpha$/Fe]
reference values. The vertical dotted line separates the RGB-a
([Fe/H]>-0.75) from the RGB-MInt and RGB-MP populations. The UVES
measurements ({\it filled circles}) are compared with the measurements
by Norris \& Da Costa (1995) and Smith et al. (2000) ({\it small empty
circles}).}

\label{alpha}
\end{figure}

We obtained a single [$\alpha$/Fe] abundance by averaging Mg, Si, Ca
and Ti. While the abundance measurements of Si, Ca and Ti are quite
safe since they are based on a few tens of lines each, the ones for Mg
are slighly more uncertain being based on less than ten lines. For
Oxygen, the situation is even more delicate since the only available
lines are the forbidden [O~I] lines at 6300~\AA\ and 6363~\AA, both
falling inside one TiO band that is most disturbing for the coolest
stars ($T_{eff}<4000^oK$). The $\sim6300$\AA\ O$_2$ telluric band falls
also in that region since the radial velocitiy of $\omega$~Cen stars is
around $\sim250$~km~s$^{-1}$. As a result, only a sub-sample of the
stars anlyzed so far has reliable O abundances. 

Due to these problems, we decided to exclude Oxygen from the estimate
of the overall $\alpha$-enhancement presented in Figure~\ref{alpha},
where our measurements are compared with the results published by
Norris \& Da Costa (1995) and Smith et al. (2000). The first thing to
note is that, even if we did not perform a detailed star-by-star
comparison, our measurements in the region $-1.5<$[Fe/H]$<-0.8$ (the
RGB-MInt) appear in reasonable agreement with the literature values.  

Another interesting effect concerns the RGB-MP stars with
$-2.0<$[Fe/H]$<-1.5$, which appear to have a slightly lower
$\alpha$-enhancement with respect to the RGB-MInt stars {\it measured
from the same authors}. As recently discussed by Shetrone et al.
(2003), a low and constant $\alpha$-enhancement, in a metallicity
regime where SNe~Ia did not play any r\^ole, indicates that any
enrichment due to SNe~II must have been driven preferentially by lower
mass SNe~II, i.e., with  $M<20_{\odot}$. A more detailed study of the
single $\alpha$-elements, on a large sample of giants of all
metallicities is thus urgently needed (see also the contribution by
Verne Smith, in these proceedings).

However, the most interesting result obtained here is that the RGB-a
stars have, on average, a lower enhancement with respect to the
RGB-MInt ones, confirming the results of Pancino et al. (2002). Thus,
if these stars formed from the $\alpha$-enriched ejecta of the previous
generations of stars in $\omega$~Cen (like the RGB-MInt, for example)
then {\it the only way to lower the $\alpha$-enhancement would be a
further enrichment by SNe~Ia}. 

In order to firmly disentangle contributions of SNe~II from SNe~Ia, it
is necessary to identify appropriate {\it tracers}, i.e., elements
that, unlike the $\alpha$-elements and the iron-peak elements, are
produced {\it almost exclusively} ($\sim$100\%) by only one type of
SNe. For example, while Eu appears as a good tracer of SNe~II
enrichment, Cu and Mn appear as promising tracers of SNe~Ia, although
their effective production mechanisms are still quite debated. To
derive accurate abundances of Eu, Cu and Mn, a full spectral synthesis
of the line profile is required, since they possess hyperfine
structure. Lines of these elements are present in our UVES spectra, so
their analysis is the next urgent task.

\begin{figure}
\centering
%\plotone{pancino.fig04.ps}
\includegraphics[width=8cm,angle=270]{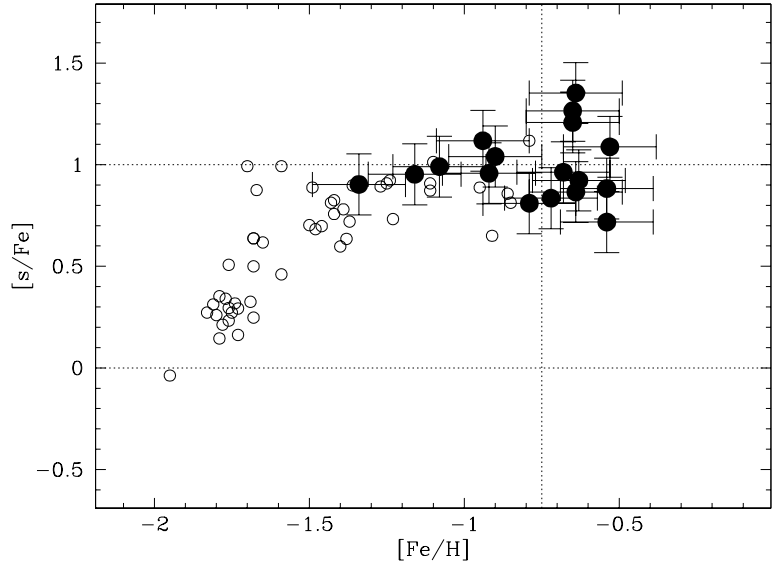}

\caption{The $s$-element abundance is plotted as a function of [Fe/H].
The horizontal dotted lines mark the solar and ten times solar [$s$/Fe]
reference values. The vertical dotted line separates the RGB-a
([Fe/H]>-0.75) from the RGB-MInt and RGB-MP populations. The UVES
measurements ({\it filled circles}) are compared with the measurements
by Norris \& Da Costa (1995), Smith et al. (2000) and Vanture et al.
(2002) ({\it small empty circles}).}

\label{esse}
\end{figure}

\subsection{The $s$-process Elements}

The next set of derived abundances concerns the $s$-process elements Y,
Zr, Ba, La, Ce and Nd, which are mainly produced in AGB stars (Busso et
al. 1999). These are some of the best studied $s$-process elements,
although their measurements for red giants are often based on a handful
of lines, so that even one single uncertain log~gf value can produce
significant variations on the final abundance. This is especially true,
in the present case, for Ce and La, that could be measured only in a
few stars, and anyway resulted in scattered abundance trends. 

To try to compensate for these drawbacks, we averaged the four most
reliable elements, Y, Zr, Ba and Nd to produce a global [$s$/Fe] ratio.
In Figure~\ref{esse}, our measurements are compared with literature
results by Norris \& Da Costa (1995), Smith et al. (2000) and Vanture
et al. (2002). While the last two studies, based on more updated atomic
data, appear in good agreement with our determinations, the data by
Norris \& Da Costa (1995) had to be shifted by $\sim0.5$~dex (a value
determined using the only star in common, ROA~371) to produce a
satisfactory agreement. This has to be ascribed entirely to the choice
of log~gfs.

Figure~\ref{esse} confirms what already found in the past for the
RGB-MP and RGB-MInt populations: a dramatic increase of the $s$-process
overabundance is clearly seen for RGB-MP stars up to approximately
[Fe/H]$\simeq-1.3$, followed by a flattening out, or at least a break
in the slope, for the RGB-MInt stars. Smith (2002) noticed that this
break in slope, if true, appears exactly at the metallicity that
separates the RGB-MP from the RGB-MInt.

Concerning the RGB-a, we see that it shows the same enrichment level of
the RGB-MInt, i.e., ten times higher than the solar value. The likely
interpretation is that the RGB-MInt and the RGB-a were similarly
enriched by low mass AGB stars ejecta. This continuity of properties
between the RGB-MInt and the RGB-a is somehow in contrast with the
discontinous behaviour of their $\alpha$-enhancement. 

\begin{figure}
\centering
%\plotone{pancino.fig05.ps}
\includegraphics[width=7cm]{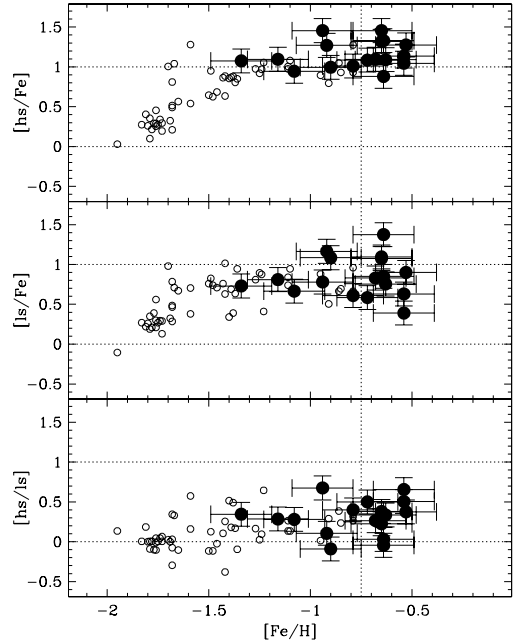}

\caption{The horizontal dotted lines mark the solar and ten times solar
[$s$/Fe] reference values, while the vertical dotted line separates the
RGB-a ([Fe/H]>-0.75) from the RGB-MInt and RGB-MP populations. UVES
measurements ({\it filled circles}) are compared with the literature
measurements as in Figure~\ref{esse} ({\it small empty circles}). The
heavy ({\it top panel}) and light ({\it middle panel}) $s$-process
elements ratios, [$hs$/Fe] and [$ls$/Fe] are plotted against [Fe/H],
together with their ratio ({\it bottom panel}) [$hs$/$ls$].}

\label{hsls}
\end{figure}

To further investigate on the source of $s$-process enrichment, we used
an indicator of the neutron exposure, or efficiency of the $s$-process:
the ratio [$hs$/$ls$] between heavy and light $s$-process elements. We
computed these ratios by averaging Ba with Nd for the heavy component
and Y with Zr for the light component. The result is shown in
Figure~\ref{hsls}, where UVES measurements are compared with literature
results, as above.

The clear overabundance of heavy $s$-sprocess elements with respect to
the light ones points towards low mass (1.5$\div$3~$M_{\odot}$) AGB
stars as the main enriching source for $\omega$~Cen, through the
$^{13}$C($\alpha$,$n$)$^{16}$O reaction that takes place in the
interpulse period (Busso et al. 1999). The RGB-a makes no exception,
showing again a perfect continuity of properties with the RGB-MInt. 

%%%%%%%%%%%%%%%%%%%%%%%%%%%%%%%%%%%%%%%%%%%%%%%%%%%%%%%%%%%%%%%%%%%%%%
%%% CONCLUSIONS %%%%%%%%%%%%%%%%%%%%%%%%%%%%%%%%%%%%%%%%%%%%%%%%%%%%%%
%%%%%%%%%%%%%%%%%%%%%%%%%%%%%%%%%%%%%%%%%%%%%%%%%%%%%%%%%%%%%%%%%%%%%%

\section{Open Problems}

The literature data from Figures~\ref{alpha} and \ref{esse} suggest
that the RGB-MP and the RGB-Mint sub-populations have been
simultaneously enriched by SNe~II and low mass AGB stars (see Norris \&
Da Costa 1995 and Smith et al. 2000). This poses a well known problem,
concerning the enrichment timescales of these two contributors. In
particular, the timescales of enrichment by SNe~II are thought to be
extremely short ($\leq1$~Gyr), so the hypothesis was made that
$\omega$~Cen reached [Fe/H]$\simeq-0.8$ in such a short time, i.e.,
soon after a burst of star formation. However, the lifetime of a
1.5$\div$3~$M_{\odot}$ AGB star is of 1--3~Gyr (Castellani, Chieffi \&
Straniero 1990), so AGB stars took {\it at least} 1--3~Gyr to enrich
stars with [Fe/H]$\leq-1.3$, where the break in the [$s$/Fe] slope with
[Fe/H] appears. 

Concerning the RGB-a, as already said, we also have some contradictory
evidence. We see a remarkable continuity of properties with the other
RGB sub-populations -- especially with the RGB-MInt -- when we look at
the enrichment in $s$-process elements, where [$s$/Fe]$\simeq+1.0$ at
all metallicities (Figure~\ref{esse}). We see instead a clearly
discontinuous behaviour in the $\alpha$-enhancement, suggesting some
additional enrichment by SNe~Ia (Figure~\ref{alpha}).

%%%%%%%%%%%%%%%%%%%%%%%%%%%%%%%%%%%%%%%%%%%%%%%%%%%%%%%%%%%%%%%%%%%%%%
%%% BIBLIOGRAFIA %%%%%%%%%%%%%%%%%%%%%%%%%%%%%%%%%%%%%%%%%%%%%%%%%%%%%
%%%%%%%%%%%%%%%%%%%%%%%%%%%%%%%%%%%%%%%%%%%%%%%%%%%%%%%%%%%%%%%%%%%%%%

\end{document}